\documentclass[prb,showpacs,preprintnumbers,floatfix,revsymb]{revtex4}
\begin{document}
\draft
\newcommand{\lsvo} {Li$_2$VOSiO$_4$ }
\newcommand{\lgvo} {Li$_2$VOGeO$_4$ }
\newcommand{\rat} {J_2/J_1 }
\newcommand{\mov} {VOMoO$_4$ }
\newcommand{\lsco} {La$_{2-x}$Sr$_x$CuO$_4$ }
\newcommand{\lscotu} {La$_{1.98}$Sr$_{0.02}$CuO$_4$ }
\newcommand{\lscofri} {La$_{1.97}$Sr$_{0.03}$CuO$_4$ }
\newcommand{\la} {$^{139}$La }
\newcommand{\cu} {$^{63}$Cu }
\newcommand{\cuo} {CuO$_2$ }
\newcommand{\ybco} {YBa$_{2}$Cu$_3$O$_{6.1}$ }
\newcommand{\ybcoca} {Y$_{1-x}$Ca$_x$Ba$_2$Cu$_3$O$_{6.1}$ }
\newcommand{\yt} {$^{89}$Y }
\newcommand{\etal} {{\it et al.} }
\newcommand{\ie} {{\it i.e.} }
\hyphenation{a-long}
%
%
%
%


\title{Frustration driven structural distortion in VOMoO$_4$
}
\author{P. Carretta\footnote{e-mail: carretta@fisicavolta.unipv.it}, N. Papinutto,
C. B. Azzoni, M. C. Mozzati and E. Pavarini} 
\address{Dipartimento di Fisica ``A. Volta'' e Unit\'a INFM di Pavia,
Via Bassi 6, 27100 Pavia, Italy} 
\author{
S. Gonthier and P. Millet}
\address{
Centre d'Elaboration des Mat\'eriaux et d'Etudes Structurales, CNRS, 31055 Toulouse Cedex, 
France
} 
\date{\today}
\widetext
\begin{abstract}





Nuclear magnetic resonance (NMR), electron paramagnetic resonance (EPR), 
magnetization measurements and electronic structure calculations
in \mov are presented. It is found that \mov is a frustrated two-dimensional antiferromagnet
on a square lattice with competing exchange interactions along the side ($J_1$) and the
diagonal ($J_2$) of the square. From magnetization measurements $J_1+J_2$ is estimated around 
$155$ K, in satisfactory agreement with the values derived from electronic structure calculations.
Around $100$ K a structural distortion, possibly driven by the frustration, is evidenced.
This distortion induces significant modifications in the NMR and
EPR spectra which  can be accounted for by valence fluctuations. The analysis of the 
spectra suggests that the size of the domains where the lattice is distorted progressively grows
on cooling as the temperature approaches the transition to the magnetic ground state at
$T_c\simeq 42$ K.    








\end{abstract}
\pacs {76.60.Es, 76.75.+i, 75.10.Jm}
\maketitle
%
\narrowtext
\section{Introduction}


In the last decade transition metal oxides have attracted a lot of interest 
in view of the rich phenomenology induced by the strong electronic correlations.
The properties of these oxides are rather peculiar once the interaction
of the electrons with the lattice becomes relevant. This is the driving mechanism
of several phenomena as, for example, superconductivity, colossal magnetoresistivity
\cite{CMR} and Spin-Peierls transition \cite{SP}. Recently, the importance of the 
coupling between the electron spin and the lattice have emerged for a new class
of materials, the frustrated antiferromagnets \cite{ST1}. In this case, the 
magnetoelastic coupling tends to relieve the degeneracy of the ground-state
caused by the frustration of
the magnetic exchange couplings \cite{Frust} . 
This is the situation observed, for example, in \lsvo \cite{lsvo1,lsvo2} which 
is a frustrated two-dimensional $S=1/2$  antiferromagnet (2DFQHAF) on a
square lattice with competing exchange interactions along the side ($J_1$) and
the diagonal ($J_2$) of the square. This compound has $J_1\simeq J_2$ ($J_1+J_2\simeq 8.5$ K)
and represents a prototype of the two-dimensional $J_1-J_2$ model, which was extensively studied
from a theoretical point of view in the last decade \cite{decade}. In the absence of any spin-lattice
coupling the ground-state is double degenerate, the two states corresponding to
collinear phases (hereafter called I and II) which differ in the orientation
of the magnetic wave-vector \cite{Chandra1}. The magnetoelastic coupling leads to a lattice distortion
at $T_{dist}\simeq (J_1+J_2)/2$,
which affects $^{29}$Si and $^7$Li NMR spectra \cite{lsvo1,lsvo2}, and \lsvo is observed to collapse 
always in one
of the two possible ground-states.


Another 2DFQHAF, nearly isostructural to \lsvo \cite{Millet}, is \mov \cite{Eick}.
The structure of these compounds is formed by pyling up
layers of SiVO$_5$ for the former and of MoVO$_5$ for the latter. These layers contain
VO$_5$ pyramids separated by (Si,Mo)O$_4$ tetrahedra (see Fig. \ref{fig:1}). The only difference
is that in \lsvo a plane of Li$^+$ ions is present between the SiVO$_5$ layers.
\mov has been recently investigated by Shiozaki and coworkers \cite{Shiozaki},
which, however, have considered it as a protoype of a weakly one-dimensional
antiferromagnet \cite{Shio2} instead of a 2DFQHAF, as it will be shown in the following.
The interest for \mov stems from the fact that although the structure is very similar
to the one of \lsvo the exchange couplings are more than an order of magnitude larger.
Thus the comparison 
of the properties of the two systems would allow to understand if frustration is indeed the
driving mechanism for the observed structural distortions.
 
 
In this manuscript nuclear magnetic resonance (NMR), electron paramagnetic resonance (EPR)
and magnetization measurements in \mov powders are presented. 
The temperature dependence of the susceptibility evidences
that also for \mov $J_1\simeq J_2$.
At temperatures below $J_1+J_2$ , namely at $T_{dist}\simeq 0.64(J_1 + J_2)$, 
\mov shows a lattice distortion possibly driven by the magnetic
frustration. The distortion seems to induce  valence fluctuations, not
observed in \lsvo , with a charge transfer from V$^{4+}$ to Mo$^{6+}$.
Moreover, it is found that domains of distorted lattice, 
with a size which progressively grows on cooling, are formed below $T_{dist}$. 
The magnitude of the superexchange couplings were estimated from electronic structure
calculations and the two-dimensional character of \mov evidenced.
Finally, the role of Mo $d$ orbitals in determining the differences with respect to \lsvo
is emphasized.


The paper is organized as follows: in Sect.II the technical aspects and the 
experimental results will be shown, while in Sect.III the analysis of the data,
including the electronic structure calculation, the analysis of NMR relaxation rates
and of the lattice distortion will be presented. The final conclusions are summarized in
Sect.IV.   





\section{Experimental aspects and experimental results}


\subsection{Sample Preparation, EPR and Magnetization Measurements}


\mov powders were obtained by solid state reaction starting from a 
stoichiometric mixture
of MoO$_3$ (Aldrich, $99.5 +$\%), V$_2$O$_5$ (Aldrich, $99.99$ \%) and V$_2$O$_3$ heated in a 
vacuum-sealed quartz
tube at 675 C for 24 hours. V$_2$O$_3$ itself was prepared by reducing V$_2$O$_5$ 
(Aldrich, $99.6 +$ \%) under
hydrogen at 800 C. The sample purity was analyzed by means of X-ray powder 
diffraction and all diffraction peaks corresponded to the ones of \mov (JCPDS 
file : 18-1454).
Single crystals were prepared by chemical transport reaction starting from 
a stoichiometric mixture
of the starting materials and TeCl4  ($10$\% in weight). The mixture was 
sealed under vacuum and heated for
24 hours at $575$ C, then slowly cooled at 10 C/hour down to room 
temperature.



EPR spectra were recorded with an X-band spectrometer equipped with a
standard microwave cavity and a variable temperature device. The measurements were performed both
on powders and on a single crystal of volume $V\leq 0.03$ mm$^3$. 
The EPR powder spectra are characterized by a lineshape
which becomes progressively
more asymmetric as the temperature is lowered 
below $130$ K (see Fig. \ref{fig:2}a and the inset to Fig. \ref{fig:3}a).
These spectra can be quite well simulated by considering a temperature independent
cylindrical $\tilde g$-tensor with 
components $g_c=1.960$ and $g_{ab}=1.932$. 
These values are consistent with the estimates of V$^{4+}$ $\tilde g$ in the framework of
the crystal field approximation, by assuming a spin-orbit coupling $\lambda = 150$ cm$^{-1}$ and
some covalency between V and O \cite{Mozza}.
It should be noticed that $g_c$ and $g_{ab}$ values are 
reversed with respect to what one might expect just by looking at the spectrum at $T=70$ K in Fig. \ref{fig:3}a.
In fact, one would be tempted to associate the low-field most intense peak with V$^{4+}$
in grains where $\vec H\perp \vec c$, while the less intense one with those grains
with $\vec H\parallel \vec c$. However, if this assignment is made the
data cannot be fitted adequately. The increase in the intensity
of the low-field peak with respect to the high-field one, below $130$ K (see the inset to
Fig. \ref{fig:3}),
has rather to be associated with a faster decrease of the linewidth $\Delta H$ 
for $\vec H\parallel \vec c$ than for $\vec H\perp \vec c$ (see Fig. \ref{fig:3}).
These results can be suitably compared to the ones derived from EPR measurements
on a single crystal. The crystal was
mounted on a sample holder that allowed to rotate the field in the $ac$ plane.
$V^{4+}$ EPR spectra on the crystal confirmed that the $\tilde g$-tensor is practically
temperature independent down to $T_c\simeq 42$ K with  
$g_c$ and $g_{ab}$
identical to the ones derived from the EPR powder spectra. Moreover, the temperature dependence 
of $\Delta H$ is the same found for the powders, 
characterized first
by a decrease on cooling, then a minimum around $60$ K, and finally an increase as the
temperature approaches $T_c$ (see Fig. \ref{fig:3}).
The temperature
dependence of the area of the EPR powder spectra, which in principle
is proportional to the static uniform
susceptibility $\chi$, is shown in Fig. \ref{fig:4}. Above $100$ K the temperature
dependence is very similar to the one derived for $\chi$ from magnetization measurements (see paragraph 
below),
however, below this temperature a rapid decrease of the EPR intensity is evident (see Fig. \ref{fig:2}b),
down to $T_c$. Then, below $T_c$ a small signal with a different $\tilde g$, possibly arising from
impurities, starts to be detected.







Magnetization ($M$) measurements were performed on \mov powders using a commercial Quantum Design
MPMS-XL7 SQUID magnetometer. The temperature dependence of the susceptibility, defined as $\chi= M/H$
with $H$ the intensity of the applied magnetic field, is shown in Fig. \ref{fig:5}a. One observes a high 
temperature Curie-Weiss behaviour, a broad maximum around $100$ K typical of low-dimensional
antiferromagnets and a kink at $T_c\simeq 42$ K, which indicates the presence of a phase transition.
This trend is the same already observed by Shiozaki et al. (Ref.  \onlinecite{Shiozaki}). At temperatures well
above the maximum the susceptibility is given by
\begin{equation}
\chi (T)={C\over T + \Theta} + \chi_{VV} ,
\label{Eq:1}
\end{equation}
where C is Curie constant, $\Theta$ the Curie-Weiss temperature and $\chi_{VV}$ Van-Vleck susceptibility.
In order to estimate $\Theta$, which for a 2DFQHAF on a square lattice is equal to $J_1 + J_2$, one 
has to determine first the value of $\chi_{VV}$. Since $\chi_{VV}$ does not
contribute to the EPR signal its value can be directly determined by plotting $\chi$ measured with
the SQUID against the EPR area for $T> 150$ K (see Fig. \ref{fig:5}b). One finds $\chi_{VV}=3.5\times 10^{-4}$ emu/mole,
a value consistent with the separation between the $t_{2g}$ levels derived from crystal field calculations
and close to the one estimated for \lsvo , where $V^{4+}$ has practically the same coordination.
Then, by fitting the susceptibility data for $T> 150$ K, with Eq. (\ref{Eq:1}) one 
derives $\Theta= 155\pm 20$ K.


\subsection{NMR spectra and relaxation rates} 


$^{95}$Mo NMR spectra were recorded both on unoriented as well as on magnetically
aligned powders by summing the Fourier transform of half
of the echo signal recorded at different frequencies. 
The powders were oriented in epoxy resin with the magnetic field direction 
along the $c$ axis which,
as can be seen from a close inspection of the crystal symmetries, corresponds to
the principal axis of the electric field gradient (EFG) at $^{95}$Mo nuclei.
The NMR spectra in the oriented powders are characterized by five
well-defined peaks (see Fig. \ref{fig:6}a) separated by
$\nu_Q\simeq 106$ kHz. For a cylindrical EFG tensor, as the one of $^{95}$Mo in \mov ,
one has that \cite{Abragam} $\nu_Q=3eV_{zz}Q(1-\gamma_{\infty})/
20 h$, where $V_{zz}$ is the principal component of the 
EFG tensor, $Q=-0.019$ barn 
$^{95}$Mo electric quadrupole moment and $(1-\gamma_{\infty})$ Sternheimer antishielding
factor. One can compare the experimental value of $\nu_Q$ with the
one derived from an estimate of the EFG based on lattice sums, 
within a point charge approximation, and
taking $(1-\gamma_{\infty})\simeq 24.861$, as estimated theoretically  \cite{Sen}. One obtains
$\nu_Q\simeq 102.5$ kHz, in remarkable agreement with the experimental finding.
On the other hand, the powder spectra are characterized by a sharp central peak, corresponding
to the $1/2\rightarrow -1/2$ transition, and by an underlying broad powder spectrum. 
Below $T_c$ $^{95}$Mo NMR powder spectrum  broadens, as expected in presence of a magnetic order yielding a local field
at the nuclei which is randomly oriented with respect to the external field.


It is interesting to analyze the temperature dependence of the shift of the 
central line in the oriented powders, for $\vec H \parallel c$,
and for the unoriented powders, which probe mainly the shift for $\vec H\perp c$. 
As reported in Fig. \ref{fig:7}a one observes that the resonance frequency of the peak 
in the unoriented powders has a 
temperature dependence which is exactly the opposite of the one observed for the susceptibility
(see Fig. \ref{fig:5}).
Since the quadrupolar corrections to the central line shift are negligible one can assert that the opposite
behaviour of these two quantities is due to a negative hyperfine coupling constant ($A$) 
between $^{95}$Mo nucleus
and the 4 nearest neighbour V$^{4+}$ ions. In fact, the shift of the NMR line can be written as
\begin{equation}
\Delta K= {4 A \chi\over g\mu_B N_A} + \delta  ,
\label{Eq:3}
\end{equation}  
with $\mu_B$ the Bohr magneton and $\delta$ the chemical shift.
Hence, by plotting $\Delta K$ vs. $\chi$ (see Fig. \ref{fig:7}b) one can derive the hyperfine 
coupling constant $A$ which
for $T\leq 100$ K turns out $A_{pow}\simeq -9$ kOe. In Fig. \ref{fig:7}b one 
clearly observes that around $100$ K there is a sizeable
change of slope which has to be associated with a marked increase in the
hyperfine coupling and suggests that around $100$ K significant modifications
in the local structure around Mo$^{6+}$ are taking place. 
Above $110$ K one has $A_{pow}\simeq -2.5$ kOe
The shift measurements in the oriented powders yield quantitatively similar results
(see Fig. \ref{fig:6}b), pointing out that
the hyperfine coupling is quite isotropic. The values for the component of the hyperfine coupling
tensor for $\vec H \parallel c$ turn out, $A_c\simeq -11.5$ kOe for $T< 100$ K and  
$A_c\simeq -2.7$ kOe for $T> 110$ K. Finally, it must be mentioned that while above $100$ K
the width of the central line is temperature independent, a sizeable broadening is observed
below $100$ K, the linewidth for $\vec H \parallel c$ 
increasing from about $2.3$ kHz at $106$ K to about $6.5$ kHz at $50$ K. This fact suggests
an increasing inhomogeneity at the microscopic level.


Nuclear spin-lattice relaxation rate $1/T_1$ was measured on the central $^{95}$Mo NMR line
by means of a saturation recovery pulse sequence. The recovery law  
was found multiexponential, as expected. Now, the point is whether the relaxation process is
driven by fluctuations of the hyperfine field or of the EFG. As we shall see later on in the 
discussion 
of the experimental results, the magnetic relaxation mechanism is the dominant one
(Sect. IIIB). Then, the recovery law
for the nuclear magnetization $m_z(t)$ is


\begin{equation}
{m_z(t\rightarrow\infty)- m_z(t)\over m_z(\infty)}= {1\over 35}e^{-{t\over T_1}} + {8\over 45}e^{-{6t\over T_1}}  +
{50\over 63}e^{-{15t\over T_1}}
\label{Eq:4}
\end{equation}   


The values of $1/T_1$ derived from the fit of the experimental data with Eq. (\ref{Eq:4}) are reported in 
Fig. \ref{fig:8}.
One notices a decrease of the relaxation rate on cooling down to about 90 K, then a plateau and
a peak at $T_c$, as expected for a second order phase transition.


The decay of the echo signal after a $\pi/2- \tau - \pi$ pulse sequence was observed to be
practically exponential. The decay of the amplitude of $^{95}$Mo echo signal arises in principle from
three different contributions, namely


\begin{equation}
E(2\tau)= E(0)\biggl[ D(2\tau)\times e^{-{2\tau\over T_1^R}}\times 
e^{-\Delta\omega '^2 \tau_c^2 f(2\tau,\tau_c)}\biggr]
\label{Eq:5}
\end{equation} 
The first term $D(2\tau)$, is the decay associated with $^{95}$Mo nuclear dipole-dipole interaction. 
The second moment $M_2$ of the corresponding frequency distribution was determined on the basis 
of lattice sums taking into account the natural abundance of $^{95}$Mo \cite{Abragam}. 
It was found that  
$\sqrt{M_2}= 90$ s$^{-1}$. If one takes this value and assumes a gaussian decay, sizeable
deviations from the exponential behavior should be detected, mainly 
at low temperature. The absence of any evidence for such gaussian deviation could stem from the 
low natural abundance of $^{95}$Mo nuclei
which leads, as for diluted nuclear spins, to a dipolar contribution to the echo decay which
is neither gaussian nor exponential \cite{Abragam}. Then one has to approximate
the nuclear dipole contribution to the echo decay with a moment expansion.  $D(2\tau)$
was expanded up to the fourth moment and its expression was used to fit the data
up to   $2\tau\simeq 1/(2\sqrt{M_2})$. It must be remarked that, in anyway, above $120$ K 
the contribution
of $D(2\tau)$ to the echo decay is small with respect to the one due to the third term in Eq.  
(\ref{Eq:5}).


The second
term in Eq.  (\ref{Eq:5}) is Redfield contribution to the echo decay. When just the central transition of an $I=5/2$ nucleus
is irradiated and in the case of an isotropic spin-lattice relaxation rate one has \cite{Walst} $1/T_1^R= 9/T_1$. The third
term is the dominant one and originates from a low-frequency dynamics characterized by a correlation
time $\tau_c$ which modulates the resonance frequency of the nuclei by $\Delta\omega '$. The observation
that the echo decay is exponential implies that $\Delta\omega '\tau_c \ll 1$, so that $f=2\tau/\tau_c$ and
the last term of Eq. (\ref{Eq:5}) becomes $exp(-2\tau/T_2)$, with \cite{Abragam} 
$1/T_2= \Delta\omega '^2\tau_c$. The temperature dependence
of $1/T_2$ derived by fitting the echo decay with Eq. (\ref{Eq:5}) is shown in Fig. \ref{fig:9}.


Finally, it must be mentioned that a $^{51}$V NMR signal was detected, with a temperature
dependence of $1/T_1$ and of $1/T_2$ very similar to the ones reported in Ref. \onlinecite{NMR}. 
However,
the values of the relaxation rates are too small to be ascribed to V$^{4+}$ sites in \mov .
The comparison of $(1/T_2)^2$ with the second moment derived for $^{51}$V dipole-dipole
interaction shows that this signal must be due to
a few percent of $^{51}$V nuclei, possibly belonging to V$^{5+}$ impurities.  


\section{Analysis of the data and discussion}


\subsection{Electronic structure and superexchange couplings}


The superexchange couplings $J_1$ and $J_2$ were estimated both theoretically, starting from electronic
structure calculations, as well as experimentally from the temperature dependence of the
susceptibility.
The electronic structure of \mov 
was calculated by using the density functional theory (DFT)
in the local density approximation (LDA).
The tight binding linear muffin 
tin orbital method\cite{lmto} 
(LMTO47 Stuttgart code) was adopted together with the exchange-correlation potential of 
Perdew and Zunger\cite{Perdew} while the lattice parameters were taken from 
Ref. \onlinecite{Eick}.           


In Fig. \ref{fig:10} the electronic structure 
of \mov and the corresponding density of states (DOS), derived 
with the linear tetrahedron method, are shown. The DOS was checked to have already 
converged with a mesh of about 858 irreducible k-points. One can notice that only two relatively narrow bands, well 
separated from all the others,  
cross the Fermi level ($\epsilon_F$), which was set to zero energy. 
The density of states shows a pronounced feature 
around $\epsilon_F$, i.e. in correspondence of 
these two bands. This feature is more evident in the lower part of
Fig. \ref{fig:10} where these two
bands and the density of states (DOS) in
the energy window (-0.8,0.4) eV are reported.
In order to minimize the linearization error
in this energy window, we placed the linearization 
energies close to the Fermi level.
The eigenvectors of the two conduction bands have
mainly V d$_{xy}$ character, mixed with some O$_2$ p$_{x/y}$
and, in the case of the lower energy band, with some  
Mo d$_{xy}$  character. 
At the $\Gamma$ point the two conduction bands are,
respectively, the bonding (lower energy) and 
antibonding (higher energy) V d$_{xy}$ bands.




The LDA bands can be understood from 
a few band tight binding model, as shown in Appendix, and the
dispersion curve of the two conduction bands can be 
written  in terms of the nearest neighbors
(NN) ($t_1$) and next nearest neighbors (NNN) ($t_2$) hoppings within the [001] plane  
and of the hopping between adjacent planes ($t_\perp$)
\begin{eqnarray}
\nonumber
\epsilon(\mathbf k) & = & \epsilon_0+
2t_2[\mathrm{cos}(k_xa)+\mathrm{cos}(k_ya)]\\
&&+ 4t_1[\mathrm{cos}(k_xa/2)\mathrm{cos} (k_ya/2)]
+ 2t_\perp\mathrm{cos}(k_zc),
\label{Eq:6}
\end{eqnarray} 
Their values can be estimated from
a least square fitting of the calculated band structure.
The results are shown in Tab.  \ref{tabmo}.
The NN hopping, $t_1$, 
and the NNN hopping, $t_2$, have
two contributions of opposite sign. The first 
one originates from the hopping between V and
NN O orbitals while the second one from hoppings involving 
V and NN O and Mo orbitals.
The sign is different because while the energy of
O$p$ orbitals lies below the Fermi level the one of Mo$d$ orbitals is above
(see Appendix).
The contribution coming from Mo 
depends mainly on the energy of Mo d$_{xy}$ effective orbital, whose energy
is affected by the hopping with  V $d_{z^2}$ and, therefore, depends
on V-Mo distance. Thus, the closer is V to Mo
the higher is the energy of Mo d$_{xy}$ and the
the smaller the contribution of Mo to $t_2$ and $t_1$. 
In addition, the hopping between O$p$ and Mo d$_{xy}$ states 
tends to enhance the ratio $t_1/t_2$ (see Appendix).



The hopping integrals can now be used 
to estimate
the exchange couplings among V$^{4+}$ spins. \mov is a half filled band Hubbard insulator and,
in the limit of strong Coulomb repulsion,
the exchange couplings can be expressed
as $J_i=4 t_i^2 /(U-V_i)$.
Here $t_i$ are the NN  and NNN hoppings,
$U$ the on-site Coulomb repulsion and 
$V_i$ the inter-site Coulomb repulsion,
which is supposed to be much smaller than $U$.
It was shown\cite{held} that typical values of
$U$ for the vanadates are $U\sim 4-5$ eV.
So, by taking $U\sim 5 $eV and neglecting $V_i$
the coupling values shown in Tab. \ref{tabmo} were derived.




\begin{table}[!ht]
\begin{center}
\begin{tabular}{c|c|c|c|c|c|c|c|c}
&$t_1$ & $t_2$ & $t_\perp$ & U & $J_1$ & $J_2$ & $J_\perp$ 
& $J_1/J_2$ \\ \hline &&&&&&&& \\
lower band  &$-110$  & $-52$ & -1& 5& 110 & 22 &$<10^{-2}$& 4.5 \\
higher band &$+135$  & $-42$ & -2& 5& 154 & 16 &$<10^{-2}$& 10
\end{tabular}
\caption{Hopping integrals (in meV) for VOMoO$_4$.
The Coulomb repulsion is in eV and the exchange 
coupling costants are in K.}
\label{tabmo}
\end{center}
\end{table}  


First one notices that $J_{\perp}$ is at least four orders 
of magnitude smaller than the in-plane coupling constants, pointing out that
\mov is a 2D system and not a one-dimensional one 
as claimed by Shiozaki and coworkers \cite{Shiozaki,Shio2}.
Second, it should be observed that the value of $J_1+J_2$ ranges between $132$ and $170$ K,
in good agreement with the value $\Theta= J_1 + J_2\simeq 155$ K derived experimentally 
for the Curie-Weiss temperature. On the other hand, for both bands we find $J_1>J_2$,
with $J_1/J_2$ around $4.5$ for the lower
energy band and $10$ for the higher energy band. This result, however, seems in contrast with
the experimental findings. In fact, the temperature dependence of the susceptibility
and in particular the ratio between $\Theta$ and the temperature of the maximum in the susceptibility
are very similar to the ones of \lsvo , pointing out that also
in \mov $J_2/J_1\simeq 1$. The similarity becomes 
evident once $\chi$ is plotted as a function of $T/\Theta $ (see Fig. \ref{fig:11}).
It is interesting to observe that an analogous discrepancy between the ratio $J_2/J_1$ derived
experimentally and the one estimated from electronic structure calculation was found by Roesner
et al. \cite{Picket} for \lsvo . Also in that case the estimate of $J_1 + J_2$ was in good agreement with the
experimental one, while the value of $J_2$ was found about a factor 10 larger than
$J_1$, the opposite of what happens for \mov . To assure that the estimate of $J_2/J_1$
was not influenced by the method adopted for calculating the band structure ,
the coupling constants were calculated also for \lsvo and a ratio $J_2/J_1\simeq 10$
was found, in good agreement 
with Roesner et al. \cite{Picket} results. 
The big difference in the ratio calculated for \lsvo and \mov cannot be associated with 
a difference in the V-O distances, which are quite similar in
both compounds, or with the small rotation of the 
basis of the  VO$_5$ pyramid. This difference should rather be ascribed to the role of Mo d$_{xy}$ orbitals in VOMoO$_4$
and of Li $s$ orbitals in Li$_2$VOSiO$_4$. As already mentioned 
the hopping between O $p$ - Mo d$_{xy}$ tends to enhance
the ratio $J_1/J_2$. On the other hand,
in \lsvo the hopping between O $p$ and NN Li $s$ orbitals 
gives a contribution to $t_1$ only, which has a sign opposite to the one due to the 
V$d$-O$p$ hopping. 
Hence the hopping through Li $s$ orbitals reduces $t_1$ and the ratio $J_1/J_2$.
 
 
Therefore, the observed discrepancies between the experimental and calculated values
of $J_2/J_1$ cannot originate from the method adopted to calculate the band 
structure but must have another origin. They should rather 
be associated with the simplified expression used to derive 
the superexchange couplings, where  just the on-site repulsion $U$ was considered.




\subsection{$^{95}$Mo relaxation rates and EPR linewidth}


As shown in the previous section $^{95}$Mo echo
decay, which probes the very low-frequency dynamics, is characterized by two regimes: 
a high temperature one where $1/T_2= \Delta\omega '^2\tau_c$ decreases
on cooling and a low temperature one where $1/T_2$ increases on approaching $T_c$ from above. 
This
means that the correlation
time $\tau_c$ which describes the dynamics decreases on cooling from room temperature 
down to $T\simeq 100$ K. 
Which could be the origin of these dynamics? One possibility is that
$^{95}$Mo echo decay above $100$ K is driven by the relaxation of unlike spins, namely, taking into account
the natural abundance and the magnitude of the nuclear magnetic moments present in \mov , of $^{51}$V spins.
Then, $\tau_c\equiv T_1$ of $^{51}$V and $\Delta\omega '$ corresponds to the nuclear dipole coupling between $^{95}$Mo
and $^{51}$V nuclei which, from lattice sums, turns out $\Delta\omega '\simeq 1510$ s$^{-1}$.  
Now
one can directly estimate $^{51}$V $1/T_1$ from $^{95}$Mo $1/T_2$ experimental data (see Fig. \ref{fig:12}) \cite{Recchia,Walst}. 


One observes  $^{51}$V $1/T_1$ increasing exponentially on decreasing temperature, 
as
one would expect for a correlated 2DQHAF \cite{Sala}. In fact, the nuclear spin-lattice relaxation rate
can be written as
\begin{equation}
{1\over T_1}= {\gamma^2\over 2N}\sum_{\vec q}\vert A(\vec q)\vert^2_{\perp} S_{\alpha\alpha}(\vec q ,\omega_L)
, \hskip 0.5 cm \alpha=x,y
\label{Eq:7}
\end{equation}   
with $\vert A(\vec q)\vert_{\perp}^2$ the form factor, which gives the hyperfine coupling of the nuclei with the
spin excitations at wave-vector $\vec q$, and $S_{\alpha\alpha}(\vec q ,\omega_L)$ the component of dynamical structure factor at nuclear
Larmor frequency. If scaling arguments apply one can express $S_{\alpha\alpha}(\vec q ,\omega_L)$ in terms of
the in-plane correlation length and, provided that \mov is in the
renormalized classical regime and 
$^{51}$V hyperfine coupling is mainly on-site, one finds
\cite{Sala}
\begin{equation}
{1\over T_1}(T)\propto \xi(T)\simeq 0.49\times exp(2\pi\rho_s/T) \biggl[ 1 -{1\over 2}({T\over 2\pi\rho_s})\biggr] ,
\label{Eq:8}
\end{equation}   
with $\rho_s$ the spin-stiffness. If $^{51}$V $1/T_1$ data are fitted with this simple 
expression a poor fitting is obtained. The point is that most of the data obtained for
$^{51}$V $1/T_1$ lie in a temperature range where $T\geq J_1+J_2\simeq 155$ K and
scaling arguments can no longer  be applied.
A more accurate quantitative analysis can be performed for 
$T\gg J_1+J_2$, where V$^{4+}$ spins are uncorrelated. In this temperature limit one can write
\cite{JPB}
\begin{equation}
{1\over T_1}= {\gamma^2\over 2}{S(S+1)\over 3}A_{\perp}^2{\sqrt{2\pi}\over\omega_E}
\label{Eq:9}
\end{equation}    
with $A_{\perp}$ $^{51}$V hyperfine coupling constants and
$\omega_E=\sqrt{J_1^2+J_2^2}(k_B/\hbar)\sqrt{2 z S(S+1)/ 3}$ the Heisenberg exchange frequency, where $z=4$ is
the number of V$^{4+}$ coupled through $J_1$ or through $J_2$, to a reference V$^{4+}$ ion. If one takes $1/T_1\simeq 6$
ms$^{-1}$ for  $T\gg J_1 + J_2$ (see Fig. \ref{fig:12}), 
one  derives $A_{\perp}\simeq 80$ kGauss. This is a typical
value for V$^{4+}$ hyperfine coupling \cite{Borsa}, supporting  the 
assumption that vanadium nuclear spin-lattice relaxation is driving $^{95}$Mo echo decay.




The increase in $^{95}$Mo $1/T_2$ on approaching T$_c$ must have a different origin since $^{51}$V $1/T_1$ is
expected to continue increasing on cooling and finally diverge at the transition temperature. The 
change of behaviour around $100$ K could be ascribed to the onset of a very low-frequency dynamics, which is possibly
associated with the motions of domain walls separating collinear I and II domains, as  recently observed in
\lsvo \cite{lsvo3}.


It is interesting to compare the temperature dependence of $^{51}$V and $^{95}$Mo nuclear 
spin-lattice
relaxation rates (see Figs. 8 and 12). One observes that while the former increases on cooling the latter
decreases. One could then be tempted to associate $^{95}$Mo relaxation to another mechanism, for example
a quadrupolar one, where the relaxation is due to phonons \cite{Abragam}. However, the nuclear spin-lattice relaxation rate due to phonons
turns out to be an order of magnitude smaller than the one derived experimentally if the recovery 
laws appropriate for a quadrupolar relaxation mechanism are used
\cite{Quadrup}. On the other hand, if one
estimates the value expected for $1/T_1$ in the assumption of a relaxation mechanism driven by 
$V^{4+}$ dynamics for $T\gg J_1+J_2$ (Eq. (\ref{Eq:9})), the calculated value turns out to be slightly larger than the experimental
one. So, it is possible that the in-plane spin correlation causes a decrease of $^{95}$Mo $1/T_1$. This is what is
expected if $^{95}$Mo form factor filters out, at least partially, the spin fluctuations at the critical
wave-vector. In fact $^{95}$Mo form factor, 
$\vert A(\vec q)\vert^2=[2A(cos(q_xa/2)+cos(q_ya/2))]^2$,
is peaked at $(q_x=0,q_y=0)$, zero at $(\pi/a,\pi/a)$ and 
reaches a reduced value at $(\pi/a,0)$ (or $(0,\pi/a)$), which
corresponds to the critical wave-vector of the envisaged collinear ground-state.   
This situation is very similar to the one found in CFTD, a non-frustrated 
2DQHAF \cite{CFTD}. In this system $^1$H have a form factor similar to the one of $^{95}$Mo 
in \mov and $1/T_1$ was also observed first to decrease on cooling 
for $T\leq J$ and
then to increase. An accurate calculation of the temperature dependence
of $^{95}$Mo $1/T_1$ in \mov  goes beyond the aim of this work, 
since it would require the precise knowledge of the 
temperature dependence of the hyperfine coupling constants between $220$ K and $T_c$.


It is also instructive to compare $^{95}$Mo $1/T_1$ with the EPR linewidth $\Delta H$ 
(see Fig. \ref{fig:3}). The similarity in the temperature dependence of both quantities is striking.
Although it is not straightforward to establish a relationship between these two 
quantities, the former probing the spectral density of the 2 spins correlation function
while the latter of the 4 spins correlation function \cite{JPB}, the physical origin
of their behavior is the same. In fact, also the initial decrease of $\Delta H$ on cooling has to be 
associated with the loss of weight of the $q\rightarrow 0$ diffusive modes and to an 
increase in the spectral weight at $(\pi/a,0)$ (or $(0,\pi/a)$), which finally gives rise,
in view of the slowing down of the critical fluctuations, to a peak at $T_c$ \cite{Ric}.
As pointed out by Richards and Salamon (Ref. \onlinecite{Ric}) the transfer of spectral weight from
$q\simeq 0$ to the critical wave vector causes also a modification in the angular dependence
of $\Delta H$, with first a decrease of $\Delta H_{ab}/\Delta H_c$ and then an increase,
exactly as it was found for \mov (see the inset to Fig. 3).
For $T\ll J_1+J_2$, the EPR linewidth should scale with the in-plane
correlation length  and if the same scaling laws derived for two-dimensional antiferromagnets \cite{JPB} 
are used,
one should find $\Delta H\propto \xi^3$. Then, by fitting the few experimental data in Fig. \ref{fig:3}
for $T\leq 55$ K and assuming the temperature dependence of $\xi$ given by Eq. (\ref{Eq:8}) one derives
a value for $2\pi\rho_s$ around $60$ K, well below $J_1+J_2$, as expected for a frustrated 2D
antiferromagnet \cite{Schulz}.




\subsection{Frustration driven structural distortion}


The analysis of NMR spectra points out
that a local structural distortion around $^{95}$Mo nuclei
takes place at  $T_{dist} \simeq 100$ K, yielding a 
sizeable change of the magnetic hyperfine
coupling (see Sect. IIB). 
The occurrence of a structural
distortion in \mov is a natural consequence of the frustration which, in the absence of spin-lattice 
interaction, for $J_2/J_1\simeq 1$ would lead to a double degenerate ground state down 
to a  temperature
 where an Ising transition to one of the two ground states occurs \cite{Chandra}.
The effect of the lattice is to relieve the degeneracy among the two ground states, so that the frustrated
system always collapses in one of the two states. This is somewhat analogous to the Jahn-Teller distortion
which relieves the degeneracy among the electronic levels split by the crystal field and for this reason
some authors have called this distortion the "Spin-Teller" distortion \cite{ST1,ST2}. 
Recently, evidence for such a distortion in a three dimensional pyrochlore antiferromagnet
was presented \cite{Keren}.
On the other hand, some connection with the
Spin-Peierls distortion is also present. In fact, quite recently Becca and Mila \cite{Becca} have shown that for a 2D $J_1-J_2$ system
the magnetoelastic coupling would induce a distortion which, depending on the value of $J_2/J_1$ and on their
dependence on the lattice parameters, could break either just the rotational invariance or, as for a standard
Spin-Peierls transition, also the translational invariance. Recent NMR measurements suggest 
that the breakdown of the rotational
invariance takes place in \lsvo \cite{Papin}.
In \mov , however, the situation is somewhat more complicated than
the one described by Becca and Mila \cite{Becca} since $J_1$ and $J_2$ show a subtle
dependence on Mo position and, therefore, cannot be expressed in a simple form in terms
of $V-V$ distance.


Now, based on simple order of magnitude estimates one can show 
that the the origin of the lattice distortion in \lsvo and \mov is the same. In fact, if one
takes the ratio between the temperature at which the distortion sets in and $J_1+J_2$, one
finds $T_{dist}/(J_1+J_2)= 0.5\pm 0.07$ for \lsvo and a close value, 
$0.64 \pm 0.07$ for \mov . This similarity can be understood by considering the expansion of 
the elastic and magnetic energies  to lowest order in the
displacements 
\begin{equation}
E=\sum_{i,j,\alpha}(\partial J_{ij}/\partial x_{\alpha})(\vec S_i .\vec S_j) x_{\alpha} +
\sum_{\alpha,\beta} k_{\alpha\beta}x_{\alpha}x_{\beta}/2
\label{Eq:10}
\end{equation}
with $x_{\alpha,\beta}$ the coordinates of the magnetic ions coupled by an elastic constant
$k_{\alpha,\beta}$. Since the reduction of magnetic energy is linear in $x_{\alpha}$ and
the elastic one is quadratic, a minimum of magnetoelastic energy can be achieved for
a small displacement $x_{eq}$ of the coordinates. Now, if one considers just $J_1$ and $J_2$
couplings the order of magnitude of the energy gain induced by the displacement turns out
to be $E_{x_{eq}}\simeq - C [\partial (J_1 + J_2)/\partial x_{\alpha}]_{x_{eq}}^2/ k_{x_{eq}}$, 
with $C$ a
constant which depends on the crystal structure. Then, if one considers that the similarities
in  \lsvo and \mov structure yield roughly similar elastic constants
and power-law dependence of $J_i$ on $x_{\alpha}$, it is likely that $k_BT_{dist}\simeq E_{x_{eq}} 
\propto (J_1+J_2)$, as experimentally
found.



It is remarkable to observe that while a clear signature of such a distortion is present 
in the NMR
spectra, no modification in $V^{4+}$ $\tilde g$-tensor is detected down to $T_c$.
On the other hand, below $100$ K a decrease in the intensity of the
EPR signal, much faster than the decrease of the macroscopic magnetization,
is observed (see Figs. \ref{fig:3} and \ref{fig:4}). This effect is not associated with a saturation or a broadening
of the EPR signal but rather indicates
that there are some ions that are becoming EPR silent. These ions
cannot correspond to V$^{4+}$, which in a pyramidal coordination as the one in \mov , should
always give an EPR signal. 
However, if valence fluctuations take place, they could correspond to Mo$^{5+}$ ions.
In fact, due to selection rules, the signal of Mo$^{5+}$ in a regular tetrahedral configuration
is cancelled out.
The regular tetrahedral coordination is indeed supported by the small
values of $^{95}$Mo quadrupolar frequency (see Sect. IIB).
As a whole, the comparison of NMR and EPR spectra leads to the following possible scenario.
The distortion induced by the frustration causes, thanks to the hybridization of Mo d-orbitals
in the band formation, a charge transfer from V$^{4+}$ to Mo$^{6+}$.
As the distortion develops it induces a modification just in the NMR spectra of 
the adjacent nuclei yielding a broadening of the NMR line (see Sect. IIB)
and the disappearence of the EPR signal of the adjacent $V^{4+}$ ions.
The $V^{4+}$ ions far from the distortion continue to give rise to an
EPR signal with unchanged $g$-values, as experimentally observed. As the temperature
is lowered the size of the distorted domains progressively grows and the EPR signal
diminshes. The formation of distorted and non-distorted domains
would support also the modifications in $^{29}$Si NMR spectra in \lsvo \cite{lsvo1}.
In fact, in \lsvo as the temperature is lowered below $T_{dist}$ one observes the 
progressive decrease of a low frequency peak (undistorted site) and the growth
of a shifted high frequency peak (distorted site) (see Ref. \onlinecite{lsvo1}).
Hence, at $T_{dist}$ a diffusive transition sets in yielding 
a progressive distortion of the whole lattice as the temperature 
decreases below $100$ K.



It should be noticed that it is somewhat unusual that Mo$^{5+}$ formation
does not cause any Jahn-Teller effect. However, in this system the Jahn-Teller distortion
could actually be hindered by the frustration driven distortion. This can occur if the distortion
yields an energy gain of the frustrated magnetic lattice larger than the shift of the
$t_{2g}$ ground-state.   
The
charge transfer could also induce a progressive crossover of \mov from
a half filled
to a quarter filled band configuration, 
as the one of NaV$_2$O$_5$ \cite{NaVO}, as the temperature decreases below $100$ K
and modifications in the transport properties should be observed. 
In fact, below $100$ K a decrease
in the energy barrier measured with resistivity is detected \cite{Transp}.




\section{Conclusions}
In conlusion, it was shown that \mov is a prototype of a 2DFQHAF on a square lattice
with $J_1\simeq J_2$, as \lsvo . The exchange couplings in \mov are much larger
than the ones of \lsvo and a value  $J_1+J_2\simeq 155$ K was derived, in good agreement with
the one estimated from electronic structure calculations. In \mov a lattice distortion
takes place at $T_{dist}\simeq 100$ K. As the temperature is lowered below $T_{dist}$
a progressive growth of the domains with lattice distortion occurs. From 
the comparison with \lsvo one finds that in these 2DFQHAF $T_{dist}$ roughly scales with $J_1+J_2$,
supporting the assumption that the distortion is driven by the magnetic frustration. 
Finally, in \mov
novel phenomena, not observed in \lsvo , occur below $T_{dist}$ and are tentatively associated 
with a charge transfer from V to Mo.


\section*{Acknowledgement}


The authors would like to thank F. Becca and A. Rigamonti for useful discussions.
The research activity in Pavia was supported by INFM project PA-MALODI. 


\appendix*



\section{Tight binding model}


The simplest model which can be used to describe the bands derived in
the framework of the LDA includes twelve orbitals.
The d$_{xy}$ of V$_1$ and $V_2$ (energy $\epsilon_d$), where
$V_1$ is the atom at $(1/4,1/4,z_V)$ and 
$V_1$ is the atom at $(-1/4,-1/4,-z_V)$; 
the 4 O p$_{x}$ orbitals (energy $\epsilon_p$)  
centered in  $(\pm (1/4-\delta_1),\mp \delta_2, \mp z_O)$, 
and $(\pm (1/4+\delta_1, \mp 1/2 \pm \delta_2, \mp z_O)$; 
the 4 O p$_y$ orbitals (energy $\epsilon_p$)
centered in
$(\mp \delta_1,\pm (1/4+\delta_2), \mp z_O)$ 
and $((\mp 1/2\pm \delta_1), \pm (1/4-\delta_2), \mp z_O)$;
the two Mo centered in $(1/4,-1/4,1/2)$ and
$(-1/4,1/4,1/2)$ (energy $\epsilon_m$).
Within this model the following hopping integrals are considered:
the hoppings between V $d$ and
O $p$ states ($t_{pd}$),  the hopping between 
NN  O p$_x$ and O $p_y$ orbitals ($t_{oo})$
and the hopping between Mo d$_{xy}$ and its NN 
O p$_x$ and O $p_y$ orbitals ($t_{mo}$).
Starting from this band model  the dispersion curve
of the two conduction band can be obtained by downfolding
all the O and V states \cite{App}. Setting $\delta_1= \delta_2= 0$ and neglecting $t_{\perp}$ one has
\begin{eqnarray}
\nonumber
\epsilon&=&\epsilon_d+8{t^2_{pd}\over \epsilon-\epsilon_p}
+4{t_{pd}^2\over \epsilon-\epsilon_p} {b\over 1-b} 
\left(\mathrm{cos}^2(k_x/2) +\mathrm{cos}^2(k_y/2)\right)
\\ &&\pm  8{t_{pd}\over \epsilon-\epsilon_p} {a\over 1-b} 
\mathrm{cos}(k_x/2)\mathrm{cos}(k_y/2) ,
\end{eqnarray}


with $$b= 16 
{t_{oo}^2 \over (\epsilon-\epsilon_p)^2}
+4{t_{mo}^2 \over (\epsilon-\epsilon_p) (\epsilon-\epsilon_m)}
\left( {1 + 4 {t_{oo} \over (\epsilon-\epsilon_p)}} \right)^2
$$
and 


$$a= 
{
4t_{oo}\over (\epsilon-\epsilon_p)}
+4{t_{mo}^2 \over (\epsilon-\epsilon_p) (\epsilon-\epsilon_m)}
\left({1 + 4 {t_{oo} \over (\epsilon-\epsilon_p)}}\right)
$$


Then one has that the in-plane NN and NNN hoppings are
$t_1=2 {t \over \epsilon -\epsilon_p} {a \over 1-b}$
and   $t_2=2 {t^2 \over \epsilon -\epsilon_p} {b \over 1-b}$,
so that $J_2/J_1\simeq (t_2/t_1)^2= b^2/a^2$.



%



\newpage



\begin{figure}
\caption{Structure of \mov projected along [001].
 VO$_5$ pyramids (black) run parallel
to the line of sight and are connected by MoO$_4$ tetrahedra (gray). For details see Ref. 
\onlinecite{Eick}. The dashed line shows the projection of the unit cell with $a= 6.6078$ \AA .}
\label{fig:1}
\end{figure}



\begin{figure}
\caption{a) Derivative of the EPR powder spectra (solid lines) at $T=293$ and $70$ K. The circles
show the best fit of the two spectra, used to determine $g$ and $\Delta H$. Notice that,
in order to better illustrate the diferrent asymmetry,
the intensity of two spectra was not normalized.
b) Derivative of the EPR powder spectra for five selected temperatures between $69$ K
and T$_c$. The temperatures, moving from the most intense to the less intense signal,
are $T= 69, 59, 49, 45.5$ and $42$ K. A marked decrease in the intensity of the EPR signal 
on cooling is evident.}
\label{fig:2}
\end{figure}



\begin{figure}
\caption{Temperature dependence of the peak to peak width of the EPR powder
spectrum in \mov . 
In the inset the ratio between the EPR linewidth for $\vec H\parallel$ and $\perp \vec c$,
estimated from the analysis of the powder spectra, is reported.}
\label{fig:3}
\end{figure}







\begin{figure}
\caption{Temperature dependence of the area of the EPR signal (squares) and of the spin
susceptibility (circles) measured experimentally with a SQUID magnetometer after subtraction of
the Van-Vleck contribution (see text). The area of the EPR signal, which in principle is
proportional to the spin susceptibility, was rescaled
to match the value of the spin susceptibility at room temperature. }
\label{fig:4}
\end{figure}




\begin{figure}
\caption{a) Temperature dependence of the susceptibility in \mov , for $H=1 $ kGauss.
The solid line shows the high temperature Curie-Weiss behaviour for $\Theta= 155$ K. b) 
Plot of the susceptibility measured with the SQUID magnetometer (after subtraction
of the atomic diamagnetic contribution)  versus the
area of the EPR signal, with the temperature as an implicit parameter. The intercept of the
solid line was used to derive 
Van-Vleck susceptibility. 
}
\label{fig:5}
\end{figure}




\begin{figure}
\caption{a) $^{95}$Mo NMR powder spectra in \mov oriented powders 
for $H= 9$ Tesla along the $c$ axis. b) $^{95}$Mo NMR shift in \mov for $\vec H\parallel c$
versus the spin susceptibility, measured with the SQUID magnetometer, after subtraction
of the Van-Vleck term. The temperature, which is an implicit parameter, is shown for a 
few selected points. The solid lines evidence the change of
slope, i.e. of hyperfine coupling, on cooling. }
\label{fig:6}
\end{figure}


\begin{figure}
\caption{a) Temperature dependence of the resonance frequency for 
$^{95}$Mo NMR central line. b) $^{95}$Mo NMR shift of the
central line in \mov unoriented powders
plotted against the the spin susceptibility, measured with the SQUID magnetometer, after subtraction
of the Van-Vleck term. The temperature, which is an implicit parameter, is shown for a 
few selected points. The solid lines evidence the change of
slope, i.e. of hyperfine coupling, on cooling.}
\label{fig:7}
\end{figure}


\begin{figure}
\caption{Temperature dependence of $^{95}$Mo NMR $1/T_1$ in \mov  for the central line
in a magnetic field of $9$ Tesla, 
derived by fitting the recovery of nuclear magnetization with Eq. (\ref{Eq:4}).}
\label{fig:8}
\end{figure}


\begin{figure}
\caption{Temperature dependence of $1/T_2$ in \mov powders for $H= 9$ Tesla, derived by fitting
the echo decay of $^{95}$Mo central transition with Eq. (\ref{Eq:5}).}
\label{fig:9}
\end{figure}


\begin{figure}
\caption{ (Top) Band structure (left) and density of states (right)
of \mov . The Fermi level
is set at zero energy. 
The symmetry points are: $\Gamma$=(0,0,0), X=($\pi/a$,0,0), 
M=($\pi/a,\pi/a$,0), Z=(0,0,$\pi/c$). (Bottom)  Band structure of \mov close to the Fermi level.
On the right side  the total DOS 
(full line) is shown. The V$d$ projected DOS
(dashed line), the O$p$ and Mo$d$ projected 
DOS (dash-dotted and dotted lines) are also shown.
}
\label{fig:10}
\end{figure}




\begin{figure}
\caption{Spin susceptibility of \lsvo and \mov as a
function of $T/\Theta$, with $\Theta= 8.7$ K and $155$ K, respectively. 
The amplitude of the susceptibility of \mov has been rescaled by a factor slightly larger
than the ratio between the Curie-Weiss temperatures indicating a slightly lower purity of
\mov sample with respect to \lsvo .}
\label{fig:11}
\end{figure}


\begin{figure}
\caption{Temperature dependence of $^{51}$V NMR $1/T_1$ as estimated from the temperature dependence
of $^{95}$Mo $1/T_2$ shown in Fig. \ref{fig:9} (see Sect. IIIB). The line is a guide to the eye.}
\label{fig:12}
\end{figure}












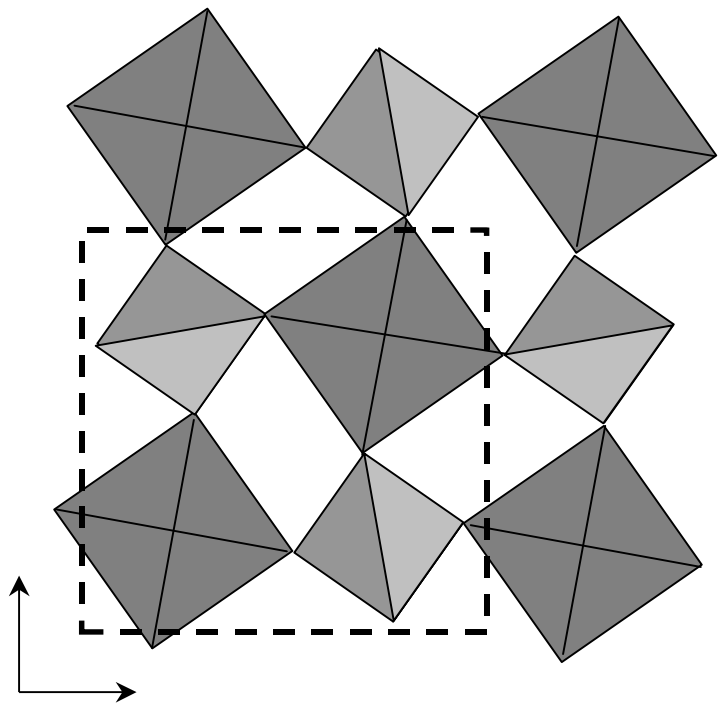
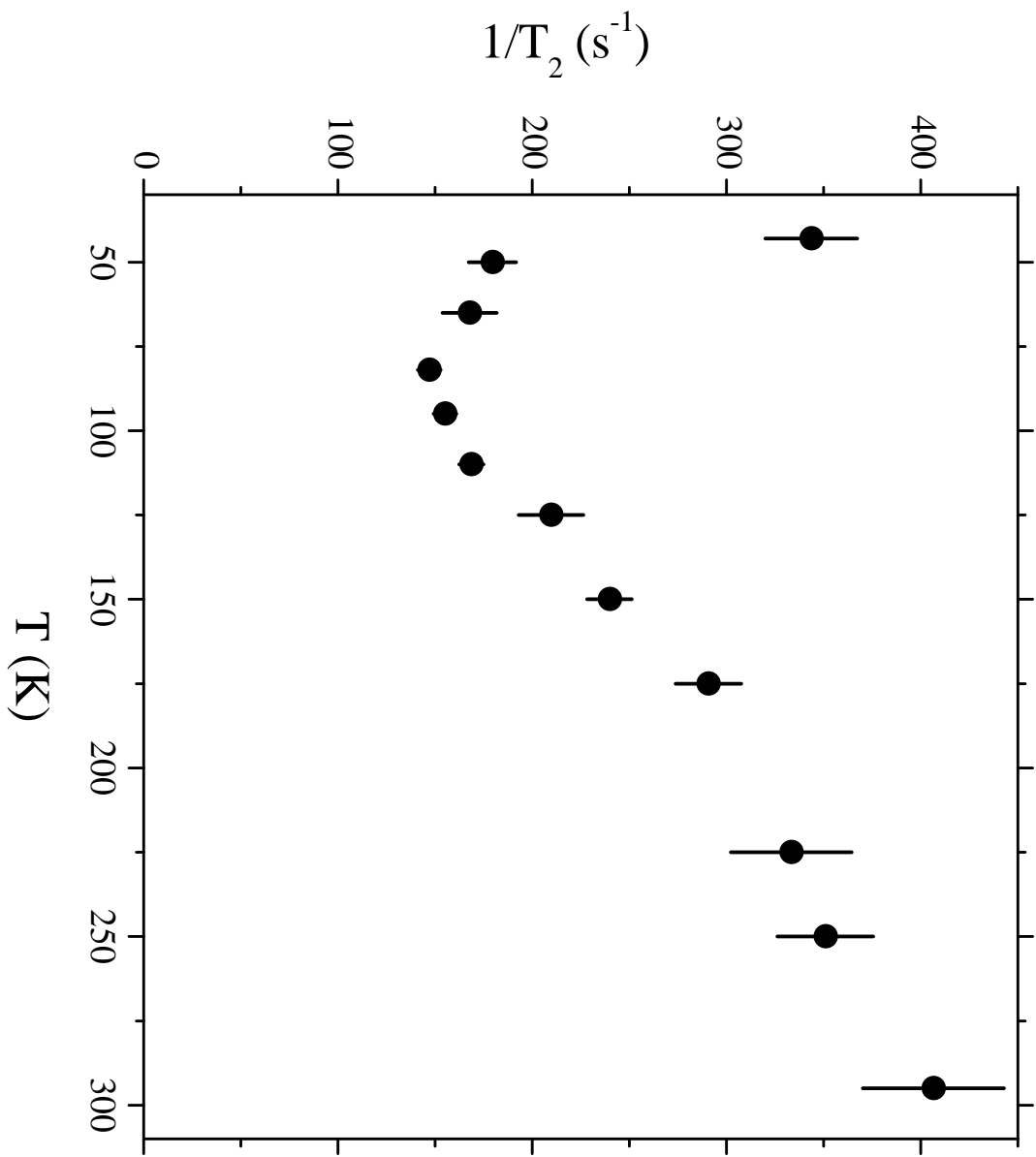
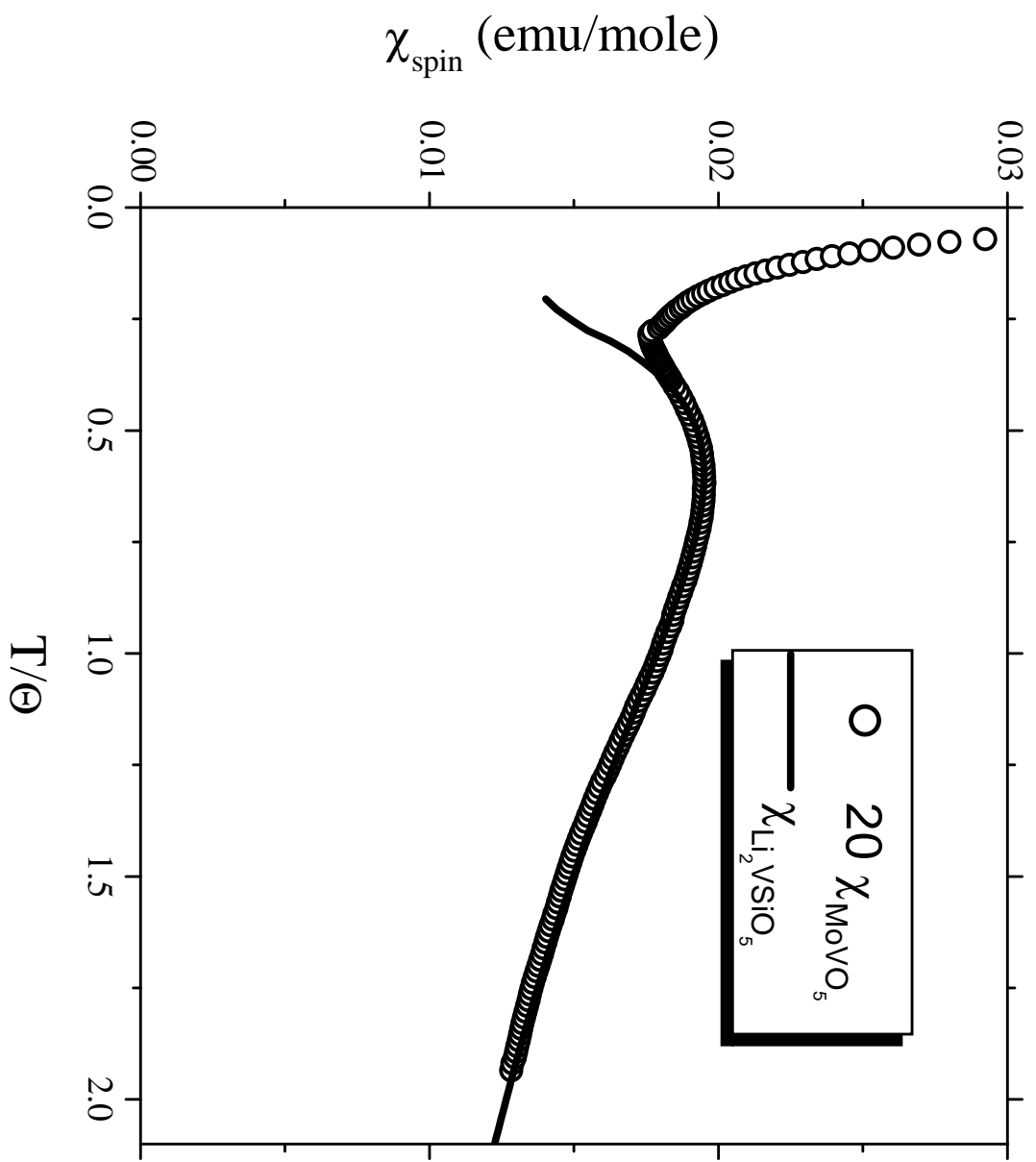
\end{document}